\documentclass[12pt, draftclsnofoot, onecolumn]{IEEEtran}

\usepackage{amsmath}
\usepackage{mathrsfs}
\usepackage{pifont}
\usepackage{bbding}
\usepackage{amsmath,epsfig}
\usepackage{stmaryrd}
\usepackage{amssymb}
\usepackage{amsfonts}
\usepackage{epic}
\usepackage{graphicx}
\usepackage{curves}
\usepackage{cite}

\usepackage{algorithm}

\usepackage{algpseudocode}
\usepackage{cases}
\usepackage{stfloats}
\usepackage{latexsym}
\usepackage{epstopdf}
\usepackage{epic}
\usepackage{bm}
\usepackage{multirow}
\usepackage{xcolor}
\usepackage{mathrsfs}
\usepackage{pifont}
\usepackage{bbding}
\usepackage{amsmath,epsfig}
\usepackage{mathbbold}
\usepackage{stmaryrd}
\usepackage{amssymb}
\usepackage{amsfonts}
\usepackage{epic}
\usepackage{graphicx}
\usepackage{curves}
\usepackage{algorithm}
\usepackage{algpseudocode}

\usepackage{stfloats}
\usepackage{latexsym}
\usepackage{epstopdf}
\usepackage{epic}
\usepackage{multirow}

\usepackage{subfigure}
\usepackage{color}

\usepackage{amsmath}
\usepackage{mathrsfs}
\usepackage{pifont}
\usepackage{bbding}
\usepackage{amsmath,epsfig}
\usepackage{mathbbold}
\usepackage{stmaryrd}
\usepackage{amssymb}
\usepackage{amsfonts}
\usepackage{epic}
\usepackage{graphicx}
\usepackage{curves}
\usepackage{cite}

\usepackage{algpseudocode}

\usepackage{stfloats}
\usepackage{latexsym}
\usepackage{epstopdf}
\usepackage{epic}
\usepackage{bm}
\usepackage{multirow}
\usepackage{xcolor}
\usepackage{subfigure}
\usepackage{array}
\usepackage{booktabs} %调整表格线与上下内容的间隔
\usepackage{multirow}
\usepackage[justification=centering]{caption}

\begin{document}
\title{Secrecy Energy Efficiency Maximization for UAV-Enabled Mobile Relaying}
\author{\IEEEauthorblockN{Lin~Xiao,~\IEEEmembership{Member,~IEEE,} Yu~Xu, Dingcheng~Yang,~\IEEEmembership{Member,~IEEE,} and~Yong~Zeng,~\IEEEmembership{Member,~IEEE}}
\thanks{The work of D.Yang was supported by National Science Foundation of China under Grant 61703197.}
\thanks{}
\thanks{}
\vspace{-4ex}
}

\maketitle

\begin{abstract}
This paper investigates the secrecy energy efficiency (SEE) maximization problem for unmanned aerial vehicle enabled mobile relaying system, where a high-mobility UAV is exploited to assist delivering the confidential information from a ground source to a legitimate ground destination with the direct link blocked, in the presence of a potential eavesdropper. We aim to maximize SEE of the UAV by jointly optimizing the communication scheduling, power allocation, and UAV trajectory over a finite time horizon. The formulated problem is a mixed-integer non-convex optimization problem that is challenging to be solved optimally. To make the problem tractable, we decompose the problem into three subproblems, and propose an efficient iterative algorithm that alternately optimizes the subproblems. In addition, two special cases are considered as benchmarks in this paper. Simulation results show that the proposed design significantly improves the SEE of the UAV, as compared to the benchmarks.
\end{abstract}

% Note that keywords are not normally used for peerreview papers.
\section{Introduction}
Recently, due to the high mobility, on-demand deployment/placement and line-of-sight (LoS) link, unmanned aerial vehicle (UAV) has attracted significant research interests in wireless communications \cite{zeng2016magazine}, such as for traffic offloading, aerial BSs, mobile relaying \cite{yang2017intelligent,chen2017uavcaching,Wu2018multiuav,Zeng2017throughput,Zhang2017mobilerelaying}, information broadcasting and data collection \cite{Zeng2018multicasting,Zhan2018energy,wu2017common,lyu2016cyclical}. Compared to the traditional terrestrial communications, UAV-enabled communications have more flexible mobility and potentially reduced cost. For one thing, UAV-enabled communication systems is especially suitable to be applied for on-demand coverage or unexpected events due to the swift and flexible deployment of UAV. For another, there is more likely to have line-of-sight (LoS) link between UAV-ground link, which can significantly improve link capacity. In addition, UAV-enabled communications provide a new degree of freedom for performance enhancement by trajectory design. Generally speaking, UAV-enabled communications can best suit the communication requirement by trajectory optimization, where the UAV is subject to practical mobility constraints, such as initial/final locations, maximum speed, and maximum acceleration. However, the limited on-board energy of a UAV is one of the biggest challenges in UAV-enabled communications since a UAV needs much additional propulsion energy to maintain aloft. As a result, the authors in \cite{Zeng2017energyefficient} obtained the analytical UAV’s energy consumption model for fixed-wing UAVAs, which was expressed as a function with respect to UAV’s speed and acceleration. Based on this, the work \cite{Zhang2017mobilerelaying} studied the spectrum and energy efficiency maximization issues in a UAV-enable communication system, in which the UAV trajectory and transmit power are jointly optimized. In particular, the UAV’s trajectory is needed to be optimized to achieve a high-rate communication with the ground nodes, while the energy consumption of the UAV is expected to be lower as much as possible \cite{Yang2018TVT}.

One particular promising use case for UAVs in wireless communications is UAV-enabled mobile relaying. Compared with traditional static relaying, UAV-enabled mobile relaying has many advantages. First, the UAV can adjust the location adaptively to suit the communication environment. Besides, the mobility of the mobile relaying offers new possibilities for on-demand communication and swift deployment/placement. In \cite{Zeng2017throughput}, the authors proposed a UAV-enabled mobile relaying system, where throughput maximization problem is considered via joint transmit power allocation and relay trajectory optimization.

On the other hand, for UAV-enabled wireless communication systems, how to ensure secure transmission of confidential information with the presence of intentional or unintentional eavesdropping is another important problem, due to the broadcast and shared nature of wireless channels. The secrecy rate is the main design metric in physical-layer security and has been investigated in many prior works (e.g., \cite{Gopala2008fading,zhao2018secureUAV,Zheng2013jamming,Li2015AFartificial,Zhang2017Globecom,Li2018wiretap},). In the existing literature on physical-layer security, one of the challenging problems is that the eavesdropper is generally passive so that it is difficult to obtain its channel state information (CSI). This motivates us to resolve the CSI of the potential eavesdropper by using UAV because the channel power gain can be easily obtained by obtaining the eavesdropper’s location, while the potential eavesdropper’s location can be detected by the UAV via a UAV-mounted camera or radar \cite{Caris2014SAR}. In \cite{Wang2017security}, the physical layer security in UAV-enabled mobile relaying system was studied with the goal of the secrecy rate (SR) maximization, but it did not focus on the UAV trajectory optimization and ignore the communication link between the source and eavesdropper. Motivated by this, we propose a new physical layer security problem in UAV-enable mobile relaying system in this paper.

\begin{figure}
\setlength{\abovecaptionskip}{0.cm}
\setlength{\belowcaptionskip}{-0.cm}
%\begin{minipage}
\centering
{\includegraphics[width=0.5\textwidth,height=2.5in]{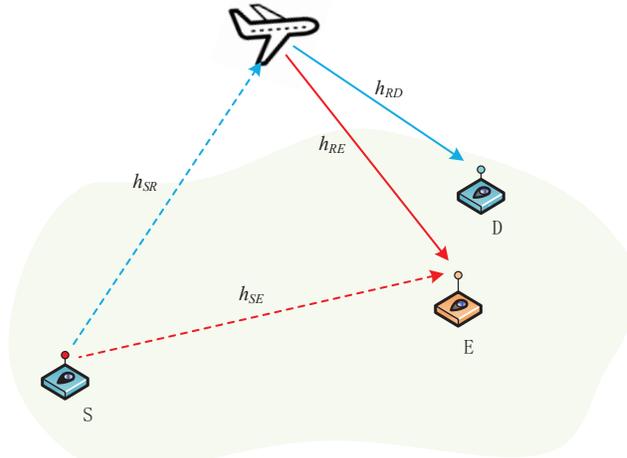}}~~~
\caption{Illustration of physical-layer security in a UAV-enabled mobile relaying system.\vspace{-3ex}} \label{model}
\end{figure}

In this paper, we consider physical-layer security in UAV-enable mobile relaying system as shown in Fig. \ref{model}, where a UAV is employed to relay the information from a ground source node to a ground destination node in the presence of a potential eavesdropper. The direct link between the source and destination is assumed to be severely blocked. Our aim is to maximize the secrecy energy efficiency (SEE) of the UAV in the finite time horizon, via jointly optimizing the communication scheduling, the source/relay power allocation, and UAV’s trajectory over a finite time horizon so as to strike a tradeoff between the secrecy rate and the energy consumption of the UAV. In our proposed design, the UAV’s mobility is subject to the initial/final location as well as the maximum speed/acceleration constraint. We assume that the UAV adopts time-division duplexing (TDD) model and decode-and-forward (DF) relaying. The formulated problem for SEE maximization is a mixed integer non-convex maximization problem that is difficult to solve optimally. To tackle this problem, we propose an efficient iterative problem by applying successive convex approximation (SCA) and Dinkelbach’s algorithm to obtain a suboptimal solution. Numerical results validate our proposed joint design method, and also show that significant performance gain is achieved, as compared to the benchmarks.

\section{System Model and Problem Formulation}
\subsection{System Model}
As shown in Fig. \ref{model}, we consider a wireless communication system where a UAV is dispatched to assist information transmission form a source node (denoted by $S$) to a legitimate destination node (denoted by $D$) in the presence of an eavesdropper (denoted by $E$). The link between $S$ and $D$ is assumed to be  severely blocked. Without loss of generality, a three-dimensional (3D) Cartesian coordinate system is considered. The nodes $S$, $D$ and $E$ are located at the fixed locations on the ground, whose horizontal coordinates are denoted by $\mathbf{w}_S=[x_s,y_s]^T,\mathbf{w}_D=[x_D,y_D]^T, \mathbf{w}_E=[x_E,x_E]^T$, respectively. The UAV is assumed to fly at a fixed altitude $H$ meters from a given initial location to a final location within a finite time horizon $T>0$. At any time instant $t\in[0,T]$, the time-varying coordinate of the UAV can be expressed as $[x(t),y(t),H]^T$, and the corresponding horizontal coordinate is denoted as $\mathbf{q}(t)=[x(t),y(t)]$. As thus, the UAV's velocity, acceleration at any time instant can be expressed as $\mathbf{v}(t)=\dot{\mathbf{q}}(t)$ and $\mathbf{a}(t)=\ddot{\mathbf{q}}(t)$ respectively.

For ease of exposition, the time horizon $T$ is discretized into $N$ time slots with a sufficiently small and equal-spaced time interval $\delta_t$, i.e. $T=N\delta_t$. For notational convenience, we let $\mathcal{N}=\{0,1,2,...n,...,N\}$ represent the time slot set. Therefore, the UAV's trajectory $\mathbf{q}(t)$ within the time interval $T$ can be approximately represented by the sequence $\{\mathbf{q}[n]=[x[n],y[n]]^T\}_{n=1}^{N}$, where $\mathbf{q}[n]\triangleq\mathbf{q}(n\delta_t)$ denotes the UAV location at time slot $n$, with $n\in \mathcal{N}$. Then, for any time slot $n$, the distance from $S$ to the mobile relaying UAV and $E$ can be respectively denoted as:
\begin{align}\label{distance_s2x}
  d_{SR}[n]&=\sqrt{\lVert\mathbf{q}[n]-\mathbf{w}_S\rVert+H^2}
\end{align}
\begin{align}\label{distance_s2e}
  d_{SE}&=\lVert\mathbf{w}_S-\mathbf{w}_E\rVert
\end{align}
Similarly, the distance from UAV to $D$ and $E$ can be respectively expressed as:
\begin{align}\label{distance_r2x}
  d_{RD}[n]&=\sqrt{\lVert\mathbf{q}[n]-\mathbf{w}_D\rVert+H^2}
\end{align}
\begin{align}\label{distance_r2e}
  d_{RE}[n]&=\sqrt{\lVert\mathbf{q}[n]-\mathbf{w}_E\rVert+H^2}
\end{align}
The initial and final locations of UAV can be expressed as $\mathbf{q}_0=[x_0,y_0]^T$ and $\mathbf{q}_F=[x_F,y_F]^T$, respectively. Hence, we have
\begin{align}\label{initial_location_constraint}
  \mathbf{q}[0]&=\mathbf{q}_0
\end{align}
\begin{align}\label{Final_location_constraint}
  \mathbf{q}[N]&=\mathbf{q}_F
\end{align}
By using Taylor expansion, the UAV's location, velocity and acceleration are related as \cite{Zeng2017energyefficient}:
\begin{align}\label{location_va_relationship}
  \mathbf{q}[n+1]&=\mathbf{q}[n]+\mathbf{v}[n]\delta_t+\frac{1}{2}\mathbf{a}[n]\delta_t^2
\end{align}
\begin{align}\label{l_velocity_a_relationship}
  \mathbf{v}[n+1]&=\mathbf{v}[n]+\mathbf{a}[n]\delta_t
\end{align}
where $n=0,1,2,3,...N-1$. We further impose the constraint that the UAV should have the same velocity at the initial and final locations, and it is subjected to the maximum velocity and acceleration. These constraints can be expressed as follows:
\begin{align}\label{va_constraints}
  \mathbf{v}[0]&=\mathbf{v}[N] \\
  \lVert\mathbf{v}[n]\rVert&\leq V_{max}, \forall n\\\label{sss}
  \lVert\mathbf{a}[n]\rVert&\leq a_{max}, \forall n
\end{align}
We assume that the ground-to-UAV/UAV-to-ground link is dominated by the LoS link. Furthermore, the Doppler effect caused by the UAV's mobility is assumed to be perfectly compensated \cite{Zeng2017energyefficient,Zeng2017throughput}. As a result, the channel power gain from $S$ to the UAV follows from the free space path loss model, which can be expressed as:
\begin{align}\label{sr_gain}
  h_{SR}[n]&=\beta_0d^{-2}_{SR}[n]=\frac{\beta_0}{\lVert\mathbf{q}[n]-\mathbf{w}_S\rVert^2+H^2}, \forall n,
\end{align}
where $\beta_0$ denotes the channel power gain at the reference distance $d_0=1$ meter. Similarly, channel power gains from the UAV to $D$ and $E$ can be respectively expressed as:
\begin{align}\label{rd_gain}
  h_{RD}[n]&=\beta_0d^{-2}_{RD}[n]=\frac{\beta_0}{\lVert\mathbf{q}[n]-\mathbf{w}_D\rVert^2+H^2}, \forall n,
\end{align}
\begin{align}\label{re_gain}
  h_{RE}[n]&=\beta_0d^{-2}_{RE}[n]=\frac{\beta_0}{\lVert\mathbf{q}[n]-\mathbf{w}_E\rVert^2+H^2}, \forall n,
\end{align}
Since the eavesdropper node and source node are located on the ground, The channel model between $S$ and $E$ is modeled to constitute both distance-dependent path loss and small-scale Rayleigh fading \cite{Feng2013Underlaying}, which can be expressed as:
\begin{align}\label{se_gain}
h_{SE}[n]&=K\zeta_{SE}[n]\beta_0d^{-\alpha}_{SE}, \forall n,
\end{align}
where $K$ is a constant determined by system parameters, $\zeta_{SE}$ denote the exponentially distributed random variable with unit mean accounting for Rayleigh fading. $\alpha$ is the path loss exponent. Denote $p_s[n]$ and $p_r[n]$ as the transmit power of the source node $S$ and the UAV, respectively. Note that $p_s[n]$ and $p_r[n]$ need to satisfy the following constraints:
\begin{align}\label{s_power}
\frac{1}{N}\sum_{n=1}^{N}p_s[n]\leq\bar{P_s}
\end{align}
\begin{align}\label{r_power}
\frac{1}{N}\sum_{n=1}^{N}p_r[n]\leq\bar{P_r}
\end{align}
where $\bar{P}_s\geq 0$ and $\bar{P}_r\geq 0$ are given average power budgets at the source node $S$ and the UAV respectively.

Besides, we assume that the half-duplex relaying with TDD is adopted by the UAV. Thus, we introduce a binary variable $\lambda[n]\in \{0,1\}$ to express the communication scheduling, with $\lambda[n]=1$, indicating that the UAV receives information from $S$, while $\lambda[n]=0$ implying that the UAV transmits information to $D$.
Therefore, at time slot $n$, the achievable rate from $S$ to the UAV can be expressed as:
\begin{align}\label{sr_rate}
r_{SR}[n]&=\lambda[n]\log_2\left(1+\frac{p_s[n]h_{SR}[n]}{\sigma^2}\right) \nonumber \\
         &=\lambda[n]\log_2\left(1+\frac{\gamma_0p_s[n]}{\lVert\mathbf{q}[n]-\mathbf{w}_S\rVert^2+H^2}\right)
\end{align}
where $\sigma^2$ is the white Gaussian noise power at the UAV receiver, and $\gamma_0=\frac{\beta_0}{\sigma^2}$ denotes the reference received signal-to-noise ratio (SNR) at the reference distance $d_0=1$ meter. The achievable rate at time slot $n$ from $S$ to $E$ can be expressed as:
\begin{align}\label{se_rate}
r_{SE}[n]&=\lambda[n]\log_2\left(1+\frac{p_s[n]h_{SE}[n]}{\sigma^2}\right) \nonumber \\
         &=\lambda[n]\log_2\left(1+\hat{h}_{SE}[n]p_s[n]\right)
\end{align}
where $\hat{h}_{SE}[n]=\frac{h_{SE}[n]}{\sigma^2}$. Similarly, the achievable rate from the mobile UAV to $D$ and $E$ in time slot $n$ can be respectively expressed as:
\begin{align}\label{rd_rate}
r_{RD}[n]&=(1-\lambda[n])\log_2\left(1+\frac{p_r[n]h_{SR}[n]}{\sigma^2}\right) \nonumber \\
         &=(1-\lambda[n])\log_2\left(1+\frac{\gamma_0p_r[n]}{\lVert\mathbf{q}[n]-\mathbf{w}_D\rVert^2+H^2}\right)
\end{align}
\begin{align}\label{re_rate}
r_{RE}[n]&=(1-\lambda[n])\log_2\left(1+\frac{p_r[n]h_{SR}[n]}{\sigma^2}\right) \nonumber \\
         &=(1-\lambda[n])\log_2\left(1+\frac{\gamma_0p_r[n]}{\lVert\mathbf{q}[n]-\mathbf{w}_E\rVert^2+H^2}\right)
\end{align}

Therefore, based on the equation (\ref{sr_rate}) and (\ref{se_rate}), the secrecy throughput from $S$ to the UAV over the $N$ time slots can be written as:
\begin{align}\label{SR_s_rate}
R_{Sec}^{sr}&=B\delta_t\sum_{n=1}^{N}\left[r_{SR}[n]-r_{SE}[n]\right]^+
\end{align}
where $[a]^+\triangleq\max(a,0)$. With (\ref{rd_rate}) and (\ref{re_rate}), the secrecy throughput from the UAV to $D$ over the $N$ time slots can be expressed as:
\begin{align}\label{RD_s_rate}
R_{Sec}^{rd}&=B\delta_t\sum_{n=1}^{N}\left[r_{RD}[n]-r_{RE}[n]\right]^+
\end{align}
We assume that the UAV adopts the DF relaying in the considered secure communications system, for which the UAV can only forward the secrecy information that has already been received from $S$. As a result, at any time slot, the total information bits that has been forwarded by the UAV  should be no more than the secrecy bits it received from $S$. Hence, by assuming that the processing delay at the UAC is one slot, the following information-causality constraints needs to be satisfied:
\begin{align}\label{casual_constraint}
r_{RD}[1]=0,& \nonumber \\
\sum_{j=2}^{n}r_{RD}[j] \leq  \sum_{j=1}^{n-1}\left[r_{SR}[j]-r_{SE}[j]\right]^+,& n=2,...,N.
\end{align}
In this paper, the communication-related energy consumption of the UAV such as signal processing is ignored as it is usually much smaller than the propulsion energy of the UAV \cite{Zeng2017energyefficient}. Based on \cite{Zeng2017energyefficient}, an effective upper bound for the propulsion energy consumption of fixed-wing UAV with velocity $\mathbf{v}[n]$ and acceleration $\mathbf{a}[n]$ can be expressed as:
\begin{align}\label{UAV_propulsion}
E_{UAV}=\delta_t\sum_{n=1}^{N}\left[c_1\lVert\mathbf{v}[n]\rVert^3+\frac{c_2}{\lVert\mathbf{v}[n]\rVert}\left(1+\frac{\lVert\mathbf{a}[n]\rVert^2}{g^2}\right)\right]+\Delta_k
\end{align}
where $c_1$ and $c_2$ are two constant parameters related to aerodynamics, $g$ represents the gravitational acceleration. Moreover, $\Delta_k=\frac{1}{2}m\left(\lVert\mathbf{v}[N]\rVert^2-\lVert\mathbf{v}[1]\rVert^2\right)$ denotes the change of kinetic energy of the UAV, whose value is only related to the UAV's mass $m$ as well as the initial and final speeds. With the constraint (\ref{va_constraints}), the kinetic energy of UAV is $\Delta_k=0$.
\subsection{Problem Formulation}
In this paper, we consider the secrecy energy efficiency (SEE) issue for the UAV-enabled mobile relaying system. Our objective is to maximize the SEE by jointly optimizing the communication scheduling, transmit power and UAV trajectory. The problem can be mathematically formulated as:
\begin{align}
(\mathbf{P}1) \max \limits_{\lambda,p_s,p_r,\mathbf{q},\mathbf{v},\mathbf{a}}&  \quad  \frac{B\sum_{n=1}^{N}\left(r_{RD}[n]-r_{RE}[n]\right)}{\sum_{n=1}^{N}\left[c_1\lVert\mathbf{v}[n]\rVert^3+\frac{c_2}{\lVert\mathbf{v}[n]\rVert}\left(1+\frac{\lVert\mathbf{a}[n]\rVert^2}{g^2}\right)\right]}\nonumber\\ s.t. \quad &\sum_{j=2}^{n}r_{RD}[j] \leq  \sum_{j=1}^{n-1}\left[r_{SR}[j]-r_{SE}[j]\right]^+, n=2,...,N.\label{casual_c}\\\label{casual_q}
& \mathbf{q}[0]=\mathbf{q}_0,\mathbf{q}[N]=\mathbf{q}_F,  \\\label{casual_lambda}
& \lambda[n]\in\{0,1\}, \\
&\mathbf{q}[n+1]=\mathbf{q}[n]+\mathbf{v}[n]\delta_t+\frac{1}{2}\mathbf{a}[n]\delta_t^2 ,\\
&\mathbf{v}[n+1]=\mathbf{v}[n]+\mathbf{a}[n]\delta_t,  \\
&\mathbf{v}[0]=\mathbf{v}[N],\\
& \lVert\mathbf{v}[n]\rVert\leq V_{max},\lVert\mathbf{a}[n]\rVert\leq a_{max}, \forall n ,\\
&\frac{1}{N}\sum_{n=1}^{N}p_s[n]\leq\bar{P_s} \\
&\frac{1}{N}\sum_{n=1}^{N}p_r[n]\leq\bar{P_r}
\end{align}
where $[\cdot]^+$s are omitted since the objective function in ($\mathbf{P}$1) and right-hand side (RHS) of (\ref{SR_s_rate}) must be non-negative at the optimal solution. Otherwise, the value of the objective function can be non-decreased by setting $p_r[n]=0$ and $p_s[n]=0$ without violating the constraints in (\ref{s_power}) and (\ref{r_power}).

Note that ($\mathbf{P}$1) is a mixed-integer non-convex problem. Firstly, the constraint (\ref{casual_c}) is non-convex, and the objective function is also non-concave and complex. Secondly, the constraint in (\ref{casual_lambda}) involves integer constraint. Therefore, the problem ($\mathbf{P}$1) is challenging to be solved optimally.

\section{Proposed Solution}
In this section, A sub-optimal solution is proposed to deal with the original problem. Successive convex approximation (SCA) and Dinkelbach's techniques are adopted to deal with the ($\mathbf{P}$1) problem, and an efficient iterative algorithm is proposed. Specifically, problem ($\mathbf{P}1$) is decomposed into three subproblems to optimize the communication scheduling variable $\{\lambda[n]\}$ and transmit power $\{p_s[n],p_r[n]\}$, as well as the UAV's trajectory $\{\mathbf{q}[n]\}$, respectively. A suboptimal solution can be obtained by alternately solving these three subproblems in an iterative manner until the algorithm converges.
\subsection{Subproblem1: Communication Scheduling Optimization}
To make the original problem more tractable, we first consider the communication scheduling issue. The binary variables $\{\lambda[n]\}$ in constraint (\ref{casual_lambda}) are proposed to relax into continuous variables: i.e.,$0\leq\lambda[n]\leq1$. Moreover, for any given transmit power $\{p_s^m[n],p_r^m[n]\}$ and UAV's trajectory $\{\mathbf{q}_m[n]\}$, where $m$ denoting the $m-th$ iteration, the energy consumption of the UAV is constant. Therefore, problem ($\mathbf{P}1$) can be equivalently reformulated as:
\begin{align}
(\mathbf{P}1.1) \nonumber \\
\max \limits_{\{\lambda[n]\}}  \quad  \sum_{n=1}^{N}\left(1-\lambda[n]\right)&\left[\log_2\left(1+\frac{\gamma_0p_r^m[n]}{\lVert\mathbf{q}_m[n]-\mathbf{w}_D\rVert^2+H^2}\right)-\log_2\left(1+\frac{\gamma_0p_r^m[n]}{\lVert\mathbf{q}_m[n]-\mathbf{w}_E\rVert^2+H^2}\right)\right]\nonumber\\
 s.t. &\quad 0\leq \lambda[n]\leq 1 \label{relax_lambda} \\
 &\sum_{j=2}^{n}\left(1-\lambda[j]\right)\log_2\left(1+\frac{\gamma_0p_r^m[j]}{\lVert\mathbf{q}_m[j]-\mathbf{w}_D\rVert^2+H^2}\right) \nonumber \\
  \leq  &\sum_{j=1}^{n-1}\lambda[j]\left(\log_2\left(1+\frac{\gamma_0p_s^m[j]}{\lVert\mathbf{q}_m[j]-\mathbf{w}_S\rVert^2+H^2}\right)-\log_2\left(1+\hat{h}_{SE}[j]p_s^m[j]\right)\right) \label{casual_c_lambda}
\end{align}
Note that the constraints in (\ref{relax_lambda}) and (\ref{casual_c_lambda}) are linear with respect to $\{\lambda[n]\}$ for given transmit power and UAV trajectory. Hence, problem ($\mathbf{P}1.1$) is a convex optimization problem that can be efficiently solved by standard convex optimization tools such as CVX \cite{Boyd2016CVX}

It is noted that in the optimal solution of the problem ($\mathbf{P}1.1$), if the optimal communication scheduling variables $\{\lambda[n]\}$ are all binary, then the relaxation is tight and the obtained solution is a locally optimal solution of the problem ($\mathbf{P}1.1$)\cite{Boyd}. Otherwise, we should reconstruct the communication scheduling solution for problem ($\mathbf{P}1.1$). To tackle this issue, it is proposed to divided each time slot into $\eta$ sub-slots. Therefore, the total available time sub-slots is $\hat{N}=\eta N, \eta \geq 1$. Then, the numbers of sub-slots assigned for the communication link from $S$ to UAV and the link from UAV to $D$ in time slot $n$ are $\Lambda_1[n] = \lfloor \eta \lambda[n] \rceil$ and $\Lambda_2[n] =\lfloor \eta - \eta \lambda[n] \rceil$ respectively,  where $\lfloor x\rceil$ indicate the nearest integer of $x$.
In order to achieve the binary solution for the communication scheduling between the link $S\rightarrow UAV$ and link $UAV \rightarrow D$, it is proposed to increase $\eta$ to guarantee $\Lambda_1[n],\Lambda_2[n]$ approach an integer. For example, suppose the optimal scheduling variable $\lambda[i]=0.71$ in slot $i$. Then, the time slot i can be divided into $\eta=100$ sub-slots, which makes $\Lambda_1[i]=\lfloor 71 \rceil =71,\Lambda_2[i]=\lfloor 29 \rceil =29$. It permits the communication scheduling solution as a binary solution with zero relaxation gap. Therefore, the binary solution for the relaxation problem ($\mathbf{P}1.1$) can be easily reconstructed by adopting the subdivision time slot method.
\subsection{Subproblem2: Transmit Power Optimization}
In this section, we investigate the transmit power optimization issue for the original problem. Considering the $(m+1)-th$ iteration, suppose the communication scheduling solution $\{\lambda_{m+1}[n]\}$ and UAV's trajectory $\{\mathbf{q}_m[n]\}$ are given. Then, the problem ($\mathbf{P}1$) can be reformulated as:
\begin{align}
(\mathbf{P}1.2)\qquad\qquad\qquad & \nonumber \\
\max \limits_{\{p_s[n],p_r[n]\}} \quad  \sum_{n=1}^{N}\left(1-\lambda_{m+1}[n]\right)& \left[\log_2\left(1+\hat{h}^m_{RD}[n]p_r[n]\right)-\log_2\left(1+\hat{h}^m_{RE}[n]p_r[n]\right)\right]\nonumber\\
 s.t. &\frac{1}{N}\sum_{n=1}^{N}p_s[n]\leq\bar{P_s} \\
&\frac{1}{N}\sum_{n=1}^{N}p_r[n]\leq\bar{P_r} \\
 \sum_{j=2}^{n}\left(1-\lambda_{m+1}[j]\right)\log_2&\left(1+\hat{h}^m_{RD}[j]p_r[j]\right) \nonumber \\
 \leq  \sum_{j=1}^{n-1}\lambda_{m+1}[j] &\left(\log_2\left(1+\hat{h}^m_{SR}[j]p_s[j]\right)-\log_2\left(1+\hat{h}_{SE}[j]p_s[j]\right)\right) \forall n. \label{casual_c_p}
\end{align}
where $\hat{h}^m_{RD}=\frac{\gamma_0}{\lVert\mathbf{q}_m[n]-\mathbf{w}_D\rVert^2+H^2}$,$\hat{h}^m_{RE}=\frac{\gamma_0}{\lVert\mathbf{q}_m[n]-\mathbf{w}_E\rVert^2+H^2}$ and $\hat{h}^m_{SR}=\frac{\gamma_0}{\lVert\mathbf{q}_m[n]-\mathbf{w}_S\rVert^2+H^2}$. Note that the problem ($\mathbf{P}1.2$) is non-convex as the objective function is non-concave and the casusality constraints in (\ref{casual_c_p}) is non-convex. To tackle this issue, convex approximation method is adopted. By introducing slack variables $\{\hat{r}_{RD}[n]\}$, we obtain the following problem:
\begin{align}
(\mathbf{P}1.2.1) \max \limits_{\{p_s[n],p_r[n]\}} \sum_{n=1}^{N}\left(1-\lambda_{m+1}[n]\right)&\left[\hat{r}_{RD}[n]-\log_2\left(1+\hat{h}^m_{RE}[n]p_r[n]\right)\right]\nonumber\\
 s.t. \frac{1}{N}\sum_{n=1}^{N}p_s[n]&\leq\bar{P_s} \\
\frac{1}{N}\sum_{n=1}^{N}p_r[n]&\leq\bar{P_r}
\end{align}
\begin{align}
 &\sum_{j=2}^{n}\left(1-\lambda_{m+1}[j]\right)\hat{r}_{RD}[j]\leq \nonumber\\
 &\qquad \qquad\sum_{j=1}^{n-1}\lambda_{m+1}[j]\left[\log_2\left(1+\hat{h}^m_{SR}[j]p_s[j]\right)-\log_2\left(1+\hat{h}_{SE}[j]p_s[j]\right)\right] \forall n. \label{casual_c_r_p}\\
 &\qquad\qquad\qquad\hat{r}_{RD}[n]\leq \log_2\left(1+\hat{h}_{RD}^m[n]p_r[n]\right), \forall n. \label{slack_r_c}
\end{align}
Specifically, it can be verified that there always exist one optimal solution to problem ($\mathbf{P}1.2.1$) satisfying the constraints in (\ref{slack_r_c}) with equalities. Otherwise, we can always decrease $p_r[n]$ without decreasing the objective function value. Therefore, problem ($\mathbf{P}1.2.1$) is equivalent to problem ($\mathbf{P}1.2$).

Note that problem ($\mathbf{P}1.2.1$) is still a non-convex problem due to the non-concave objective functions as well as the non-convex constraint in (\ref{casual_c_r_p}). To deal with the non-convexity, SCA technique is applied to obtain an approximate solution. In particular, for the term of $\log_2(1+\hat{h}_{RE}^m[n]p_r[n])$ in the objective function of ($\mathbf{P}1.2.1$), it is concave with respect to $p_r[n]$. Thus, its first-order Taylor expansion at the given local point $\{p_r^m[n]\}$ is a global over-estimator, i.e.,
\begin{align}\label{re_ub}
 \log_2\left(1+\hat{h}^m_{RE}[n]p_r[n]\right) \leq \log_2\left(1+\hat{h}_{RE}^m[n]p^m_r[n]\right)+\frac{\hat{h}^m_{RE}[n]\left(p_r[n]-p^m_r[n]\right)}{1+\hat{h}_{RE}^m[n]p_r^m[n]}\triangleq \hat{r}_{RE}^{ub}[n]
\end{align}
Similarly, for the concave term of $\log_2(1+\hat{h}_{SE}p_s[j])$ in the RHS of the constraint (\ref{casual_c_r_p}), by the given point $p_s^m[j]$, it can be upper-bounded as:
\begin{align}\label{se_ub}
 \log_2\left(1+\hat{h}_{SE}[j]p_s[j]\right) \leq \log_2\left(1+\hat{h}_{SE}[j]p^m_s[j]\right) +\frac{\hat{h}_{SE}[j]\left(p_s[j]-p^m_s[j]\right)}{1+\hat{h}_{SE}[j]p_s^m[j]}\triangleq \hat{r}_{SE}^{ub}[j]
\end{align}
Adopting (\ref{re_ub}) and (\ref{se_ub}) to reformulated the problem ($\mathbf{P}1.2.1$), it can be approximated as the following problem, which can be presented as:
\begin{align}
(\mathbf{P}1.2.2) \max \limits_{\{p_s[n],p_r[n]\}}& \sum_{n=1}^{N}\left(1-\lambda_{m+1}[n]\right)\left[\hat{r}_{RD}[n]-\hat{r}_{RE}^{ub}[n]\right]\nonumber\\
 s.t. & (\ref{s_power}),(\ref{r_power}),(\ref{slack_r_c}) \nonumber \\ \sum_{j=2}^{n}\left[\left(1-\lambda_{m+1}[j]\hat{r}_{RD}[j]\right)\right] \leq & \sum_{j=1}^{n-1} \lambda_{m+1}[j]\left[\log_2\left(1+\hat{h}_{SR}^m[j]p_s[j]\right)-\hat{r}_{SE}^{ub}[j]\right]
\end{align}
Note that the problem ($\mathbf{P}1.2.2$) is convex problem, which can be efficiently solved by standard convex optimization tools such as CVX. Moreover, the first-order Taylor expansion in (\ref{re_ub}) suggests that the objective values of ($\mathbf{P}1.2.1$) and ($\mathbf{P}1.2.2$) are equal only at $\{p_r^m[n]\}$, and problem ($\mathbf{P}1.2.2$) maximizes the lower bound of the objective function of problem ($\mathbf{P}1.2.1$), the objective value of ($\mathbf{P}1.2.1$) with the solution achieved by solving problem ($\mathbf{P}1.2.2$) is always no smaller than that with the given point $\{p_r^m[n]\}$.

\subsection{Subproblem 3: Trajectory Optimization}
In this section, we consider UAV trajectory optimization issue of the original problem ($\mathbf{P}1$) with the fixed communication scheduling $\{\lambda_{m+1}[n]\}$ and transmit power $\{p_s^{m+1}[n],p_r^{m+1}[n]\}$ at the $(m+1)$-th iteration. This subproblem is formulated as:
\begin{align}
(\mathbf{P}1.3)\max \limits_{\{\mathbf{q}[n],\mathbf{v}[n],\mathbf{a}[n]\}} &\frac{\sum_{n=1}^{N}\left(1-\lambda_{m+1}[n]\right)\left[\log_2\left(1+\frac{\gamma_0p_r^{m+1}[n]}{\lVert\mathbf{q}[n]-\mathbf{w}_D\rVert^2+H^2}\right)-\log_2\left(1+\frac{\gamma_0p_r^{m+1}[n]}{\lVert\mathbf{q}[n]-\mathbf{w}_E\rVert^2+H^2}\right)\right]}{\sum_{n=1}^{N}\left[c_1\lVert\mathbf{v}[n]\rVert^3+\frac{c_2}{\lVert\mathbf{v}[n]\rVert}\left(1+\frac{\lVert\mathbf{a}[n]\rVert^2}{g^2}\right)\right]} \nonumber \\
& \qquad\qquad \qquad s.t.(\ref{initial_location_constraint})-(\ref{sss}),(\ref{casual_c})\nonumber
%&\qquad s.t.  (\ref{initial_location_constraint})-(\ref{sss}),(\ref{casual_c)
\end{align}
Note that this subproblem is challenging to solve optimally due to the non-convexity of the objective function of ($\mathbf{P}1$) and the information-causality constraints in (\ref{casual_c}) with respect to the trajectory $\{\mathbf{q}[n]\}$. As a result, we propose an approximate solution by applying SCA method and Dinkelbach's algorithm. First, by introducing slack variables $\mathbf{U}_{rd}=\{u_{rd}[n]\}$, $\mathbf{U}_{re}=\{u_{re}[n]\}$, $\mathbf{U}_{sr}=\{u_{sr}[n]\}$, $\bm{\tau} = \{\tau[n]\}$, and $\mathbf{s}=\{s[n]\}$, the problem ($\mathbf{P}1.3$) can be reformulated as:
\begin{align}
(\mathbf{P}1.3.1) \max \limits_{\{\mathbf{q},\mathbf{v},\mathbf{a},\mathbf{U}_{rd},\mathbf{U}_{re},\mathbf{U}_{sr},\bm{\tau},\mathbf{s}\}} &\frac{\sum_{n=1}^{N}\left[\left(1-\lambda_{m+1}[n]\right)\left(s[n]-\log_2\left(1+\frac{\gamma_0p_r^{m+1}[n]}{u_{re}[n]}\right)\right)\right]}{\sum_{n=1}^{N}\left[c_1\lVert\mathbf{v}[n]\rVert^3+\frac{c_2}{\tau[n]}\left(1+\frac{\lVert\mathbf{a}[n]\rVert^2}{g^2}\right)\right]} \nonumber\\
&\qquad  s.t.  (\ref{initial_location_constraint})-(\ref{sss})\nonumber\\
&\qquad \quad u_{rd}\geq \lVert\mathbf{q}[n]-\mathbf{w}_D\rVert^2+H^2, \forall n, \label{u_rd}\\
&\qquad \quad u_{re}\leq \lVert\mathbf{q}[n]-\mathbf{w}_E\rVert^2+H^2, \forall n,\label{u_re} \\
&\qquad \quad u_{re}\geq 0, \label{u_re_0}\forall n, \\
&\qquad  \quad u_{sr}\geq \lVert\mathbf{q}[n]-\mathbf{w}_S\rVert^2+H^2, \forall n, \label{u_sr}\\
&\qquad \quad \tau^2[n]\leq \lVert\mathbf{v}[n]\rVert^2, \forall n, \label{tao}\\
&\qquad  \quad s[n]\leq \log_2\left(1+\frac{\gamma_0p_r^{m+1}[n]}{u_{rd}[n]}\right), \forall n,\label{ss}\\
\sum_{j=2}^{n}\left[(1-\lambda_{m+1}[j])s[j]\right] \leq& \sum_{j=1}^{n-1}\lambda_{m+1}[j] \left[\log_2\left(1+\frac{\gamma_0p_s^{m+1}[j]}{u_{sr}[j]}\right)-\log_2\left(1+\hat{h}_{SE}p_s^{m+1}[j]\right)\right],\forall n,\label{casual_1.3.1}\\
&\qquad \quad s[n]\geq 0, \forall n \label{s_0}
\end{align}
It can be verified that at the optimal solution of problem ($\mathbf{P}1.3.1)$), the constraints (\ref{u_rd}),(\ref{u_re}),(\ref{tao}-\ref{casual_1.3.1}) should hold with equalities. This is due to that: $u_{rd}[n]$ can be decreased to improve the objective value; $u_{re}[n],\tau[n]$ can be increased to improve the objective value; $u_{sr}[n]$ can be decreased to achieve a larger upper bound of the objective function, and $s[n]$ can be increased to achieve a larger lower upper bound of the objective function, which lead to the non-decreasing of the objective value. Therefore, problems ($\mathbf{P}1.3$) and ($\mathbf{P}1.3.1$) are equivalent. Note that the problem ($\mathbf{P}1.3.1$) is a non-convex problem since the constraints (\ref{u_re}),(\ref{tao}-\ref{casual_1.3.1}) are non-convex. Next, we focus on solving the non-convexity of problem ($\mathbf{P}1.3.1$).

As for the constraint in (\ref{u_re}), the term $\lvert\mathbf{q}[n]-\mathbf{w}_E\rVert^2+H^2$ is convex with respect to $\mathbf{q}[n]$. Hence, we can obtain its lower bound function via its first-order Taylor expansion at any given point $\{\mathbf{q}_m[n]\}$, which is a global under-estimator \cite{Boyd2016CVX},\cite{Wu2018multiuav}, i.e.,
\begin{align}\label{re_lb}
 \lVert\mathbf{q}[n]-\mathbf{w}_E\rVert^2 +H^2 &\geq \lVert\mathbf{q}_m[n]-\mathbf{w}_E\rVert^2 +H^2+2(\mathbf{q}_m[n]-\mathbf{w}_E)^T(\mathbf{q}[n]-\mathbf{q}_m[n])\triangleq \varphi_{re}^{lb}[n],
\end{align}
Similarly, for the RHS of the constraints in (\ref{tao}) and (\ref{ss}), they are convex functions of $\mathbf{v}[n]$ and $u_{rd}[n]$, respectively. Thus, based on the Taylor approximation method, we can obtain their lower bound functions by given any point $\{\mathbf{v}[n]\}$ and $\{u_{rd}^{m}[n]\}$, respectively, i.e.,
\begin{align}\label{v_lb}
 \lVert\mathbf{v}[n]\rVert^2\geq \lVert\mathbf{v}_m[n]\rVert^2 +2\left(\mathbf{v}_m[n]\right)^T(\mathbf{v}[n]-\mathbf{v}_m[n])\triangleq\varphi_v^{lb}[n],
\end{align}
\begin{align}\label{rd_lb}
 \log_2&\left(1+\frac{\gamma_0p_r^{m+1}[n]}{u_{rd}[n]}\right)\geq A_{rd}^m[n]-B_{rd}^m[n]\left(u_{rd}[n]-u_{rd}^m[n]\right) \triangleq \varphi_{rd}^{lb}[n],
\end{align}
where
\begin{align}\label{rd_A_lb}
A_{rd}^m[n] = \log_2\left(1+\frac{\gamma_0p_r^{m+1}[n]}{u_{rd}^m[n]}\right)
\end{align}
\begin{align}\label{rd_B_lb}
B_{rd}^m[n] =\frac{1}{\ln2}\frac{\gamma_0p_r^{m+1}[n]}{u_{rd}^m[n]\left(u_{rd}^m[n]+\gamma_0p_r^{m+1}[n]\right)}
\end{align}
Note that the information-causality constraint in (\ref{casual_1.3.1}) is non-convex. To make it feasible for convex optimization, SCA method is applied. Note that the term of $\log_2\left(1+\frac{\gamma_0p_s^{m+1}[j]}{u_{sr}[j]}\right)$ in (\ref{casual_1.3.1}) is a convex function with respect to $u_{sr}[j]$, which can be lower bound at any given point $\{u_{sr}^m[j]\}$ as follows:
\begin{align}\label{sr_lb}
 \log_2&\left(1+\frac{\gamma_0p_s^{m+1}[j]}{u_{sr}[j]}\right)\geq A_{sr}^m[j]-B_{sr}^m[j]\left(u_{sr}[j]-u_{sr}^m[j]\right) \triangleq \varphi_{sr}^{lb}[j],
\end{align}
where
\begin{align}\label{sr_A_lb}
A_{sr}^m[j] = \log_2\left(1+\frac{\gamma_0p_s^{m+1}[j]}{u_{sr}^m[j]}\right)
\end{align}
\begin{align}\label{sr_B_lb}
B_{sr}^m[j] =\frac{1}{\ln2}\frac{\gamma_0p_s^{m+1}[j]}{u_{sr}^m[j]\left(u_{sr}^m[j]+\gamma_0p_s^{m+1}[j]\right)}
\end{align}
Therefore, with (\ref{re_lb}),(\ref{v_lb}),(\ref{rd_lb}) and (\ref{sr_lb}), problem ($\mathbf{P}1.3.1$) is reformulated as:
\begin{align}
(\mathbf{P}1.3.2)
\max \limits_{\{\mathbf{q},\mathbf{v},\mathbf{a},\mathbf{U}_{rd},\mathbf{U}_{re},\mathbf{U}_{sr},\bm{\tau},\mathbf{s}\}} & \frac{\sum_{n=1}^{N}\left[\left(1-\lambda_{m+1}[n]\right)\left(s[n]-\log_2\left(1+\frac{\gamma_0p_r^{m+1}[n]}{u_{re}[n]}\right)\right)\right]}{\sum_{n=1}^{N}\left[c_1\lVert\mathbf{v}[n]\rVert^3+\frac{c_2}{\tau[n]}\left(1+\frac{\lVert\mathbf{a}[n]\rVert^2}{g^2}\right)\right]} \nonumber \\
&  s.t.  (\ref{initial_location_constraint})-(\ref{sss}),(\ref{u_rd}),(\ref{u_re_0}),(\ref{u_sr}),(\ref{s_0})\nonumber\\
& u_{re}\leq \varphi_{re}^{lb}[n],\quad \forall n,   \\
& \tau^2[n]\leq \varphi_v^{lb}[n], \quad\forall n, \\
& s[n]\leq \varphi_{rd}^{lb}[n], \qquad \forall n, \\
\sum_{j=2}^{n}\left[(1-\lambda_{m+1}[j])s[j]\right] \leq &\sum_{j=1}^{n-1}\lambda_{m+1}[j]\left[\varphi_{sr}^{lb}[j]-\log_2\left(1+\hat{h}_{SE}p_s^{m+1}[j]\right)\right],\forall n,
\end{align}
Note that all constraints of problem ($\mathbf{P}1.3.2$) are convex, and the objective function of ($\mathbf{P}1.3.2$) consists of a concave numerator and a convex denominator, which motives us to employ the Dinkelbach's algorithm to efficiently solve this fractional programming problem \cite{Zeng2017energyefficient,Wu2018multiuav}. In addition, the problem ($\mathbf{P}1.3.2$) actually maximizes the lower bound of the objective function, as a result, the obtained solution by solving ($\mathbf{P}1.3.1$) should be non-decreasing over iterations.

\subsection{Overall Algorithm}
\begin{table}
\centering
\begin{tabular}{lcl}
\hline
\textbf{Algorithm 1}: Proposed solution for solving problem ($\mathbf{P}1$)\\
\hline
\specialrule{0em}{3pt}{3pt}
1: Initialize $\{p_s^m[n],p_r^m[n],\mathbf{q}_m[n],\mathbf{v}_m[n]\}$, slack variables \\
 $\{u_{rd}^m[n],u_{sr}^m[n]\}$, accuracy $\epsilon \ge 0$, Let $m=0$;\\
\specialrule{0em}{3pt}{3pt}
2: \textbf{Repeat:}\\
\specialrule{0em}{3pt}{3pt}
3: Solve problem ($\mathbf{P}1.1$) for the given $\{p_s^m[n],p_r^m[n],$\\
$\mathbf{q}_m[n],\mathbf{v}_m[n]\}$, denote the optimal solution as $\{\lambda_{m+1}[n]\}$;\\
\specialrule{0em}{3pt}{3pt}
4: Solve problem ($\mathbf{P}1.2$) for the given $\{\lambda_{m+1}[n],\mathbf{q}_m[n],$\\
$\mathbf{v}_m[n]\}$, denote the optimal solution as $\{p_s^{m+1}[n],p_r^{m+1}[n]\}$;\\
\specialrule{0em}{3pt}{3pt}
5: Solve problem ($\mathbf{P}1.3.1$) for the given $\{\lambda_{m+1}[n],p_s^{m+1}[n],$\\
$p_r^{m+1}[n],\mathbf{q}_m[n],\mathbf{v}_m[n]\}$ and $\{u_{rd}^m[n],u_{sr}^m[n]\}$, and denote the\\
optimal solution as $\{\mathbf{q}_{m+1}[n],\mathbf{v}_{m+1}[n],u_{rd}^{m+1}[n],u_{sr}^{m+1}[n]\}$.\\
\specialrule{0em}{3pt}{3pt}
6: Update $m\leftarrow m+1$\\
\specialrule{0em}{3pt}{3pt}
7: \textbf{Until}: Converges to the prescribed accuracy $\epsilon$.\\
\specialrule{0em}{3pt}{3pt}
\hline
\\
\end{tabular}
\caption{The Overall Algorithm Description for Problem $\mathbf{P}1$}\label{algorithm}
\end{table}
In summary, an efficient algorithm is proposed to solve problem ($\mathbf{P}1.1$) ($\mathbf{P}1.2$) and  ($\mathbf{P}1.3.1$) alternately via applying the SCA technique and Dinkelbach's algorithm. Since the optimal value of ($\mathbf{P}1$) is finite, and the objective value of ($\mathbf{P}1$) with the solutions obtained by solving these subproblems is non-decreasing in each iterations, the proposed iterative algorithm is guaranteed to converge. The details of the proposed algorithm are shown in Algorithm 1.

\subsection{UAV Initial Trajectory}
In order to properly balance the causality constraints of the system, in this subsection, the UAV initial trajectory is set as a double-circular initial trajectory as shown in Fig.\ref{model}, where the UAV flies follow two circles. The two circles are connected by a straight line segment (from point $I$ to point $F$) as described in Fig.\ref{two_circular}. The UAV is assumed to start the task over the initial location (denoted by $I$) and flies over the first circle with one lap (It can also be set as several laps such as $\mathit{l}$ laps, if the time $N$ is sufficiently large). After that, the UAV takes straight flight over the straight line ($I \rightarrow F$). Moreover, it operate its second circular flight around the final location point $F$ for one lap (It can also be set as several laps, depending on UAV's speed and flying duration). At last, it finishes the task, landing at the final location $F$.

For simplifying the initialization procedure, suppose the speed of UAV during the period is a constant value $V, 0\leq V\leq V_{max}$. Assume that the duration for UAV flying through the straight line ($I\rightarrow F$) is denoted by $N_s$ time slots. In addition, suppose the UAV flies over the two circles with the same time slots $N_c$, thus we have $N_c=\frac{N-N_s}{2}$. Moreover, it means that the two circles have a same radius $r$, which can be denoted by $r=\frac{ VN_c\delta_t}{2\pi}$. It is noted that we need to let $N_c$ as an integer, define $N_s = f_{even}(T/V_s)$, where $f_{even}(x)$ represents that $x$ rounds to the nearest even number, such as $f_{even}(3.7)=4$, $f_{even}(8.9)=8$. Therefore, given the initial and final location coordinate $\mathbf{q}_0=[x_0,y_0]^T$ and $\mathbf{q}_F=[x_F,y_F]^T$, the two circular trajectories can be respectively expressed as:
\begin{figure}
\setlength{\abovecaptionskip}{-0.cm}
\setlength{\belowcaptionskip}{-0.cm}
%\begin{minipage}
\centering
{\includegraphics[width=0.5\textwidth,height=2.5in]{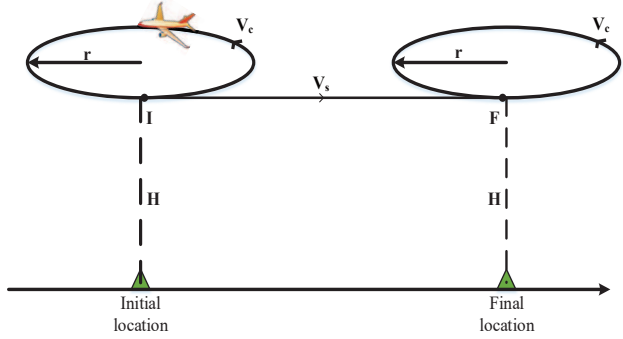}}~~~
\caption{Illustration of double-circular UAV initial trajectory.\vspace{-3ex}} \label{two_circular}
\end{figure}
\begin{align}\label{q_c1_T}
\mathbf{q}_{c1}^0[n]=&\left[x_0+r\cos\left(\frac{2\pi(n-1)}{N_c-1}-\frac{\pi}{2}\right)\right.\nonumber\\
&\qquad \quad  \left.(y_0+r)+r\sin\left(\frac{2\pi(n-1)}{N_c-1}-\frac{\pi}{2}\right)\right]\nonumber \\
& \qquad \qquad \qquad \qquad \qquad \qquad n=1,2,3,...,N_c
\end{align}
\begin{align}\label{q_c2_T}
\mathbf{q}_{c2}^0[n]=&\left[x_F+r\cos\left(\frac{2\pi(n-N_c-N_s-1)}{N_c-1}-\frac{\pi}{2}\right)\right.\nonumber\\
& \quad  \left.(y_F+r)+r\sin\left(\frac{2\pi(n-N_c-N_s-1)}{N_c-1}-\frac{\pi}{2}\right)\right]\nonumber \\
& \qquad \qquad \qquad \qquad n=N_c+N_s+1,...,N,
\end{align}
where $(x_0,y_0+r)$ and $(x_F,y_F+r)$ represent the coordinates of the centre of the two circles, respectively.

For the UAV's flight over the straight segment ($I \rightarrow F$), the UAV flies with constant speed. Thus, the trajectory from $I$ to $F$ can be expressed as:
\begin{align}\label{q_s_T}
\mathbf{q}_{s}^0[n]= \mathbf{q}_0+\frac{n-N_c}{N_s}\left(\mathbf{q}_F-\mathbf{q}_0\right), n=N_c+1,....,N_c+N_s
\end{align}
In summary, the double-circular initial trajectory of the UAV can be given as follows:
\begin{equation}
\mathbf{q}^0[n]=\left\{
             \begin{array}{lr}
             \mathbf{q}_{c1}^0[n],\qquad &n=1,2,...,N_c,   \\
             \mathbf{q}_{s}^0[n],\qquad &n=N_c+1,...,N_c+N_s,\\
             \mathbf{q}_{c2}^0[n],\qquad &n=N_c+N_s+1,...,N,
             \end{array}
\right.
\end{equation}
\begin{figure*}
\vspace{-0.3cm}
\setlength{\abovecaptionskip}{0.cm}
\setlength{\belowcaptionskip}{0.cm}
\centering
\subfigure[$T=100s$]{\includegraphics[width=0.5\textwidth]{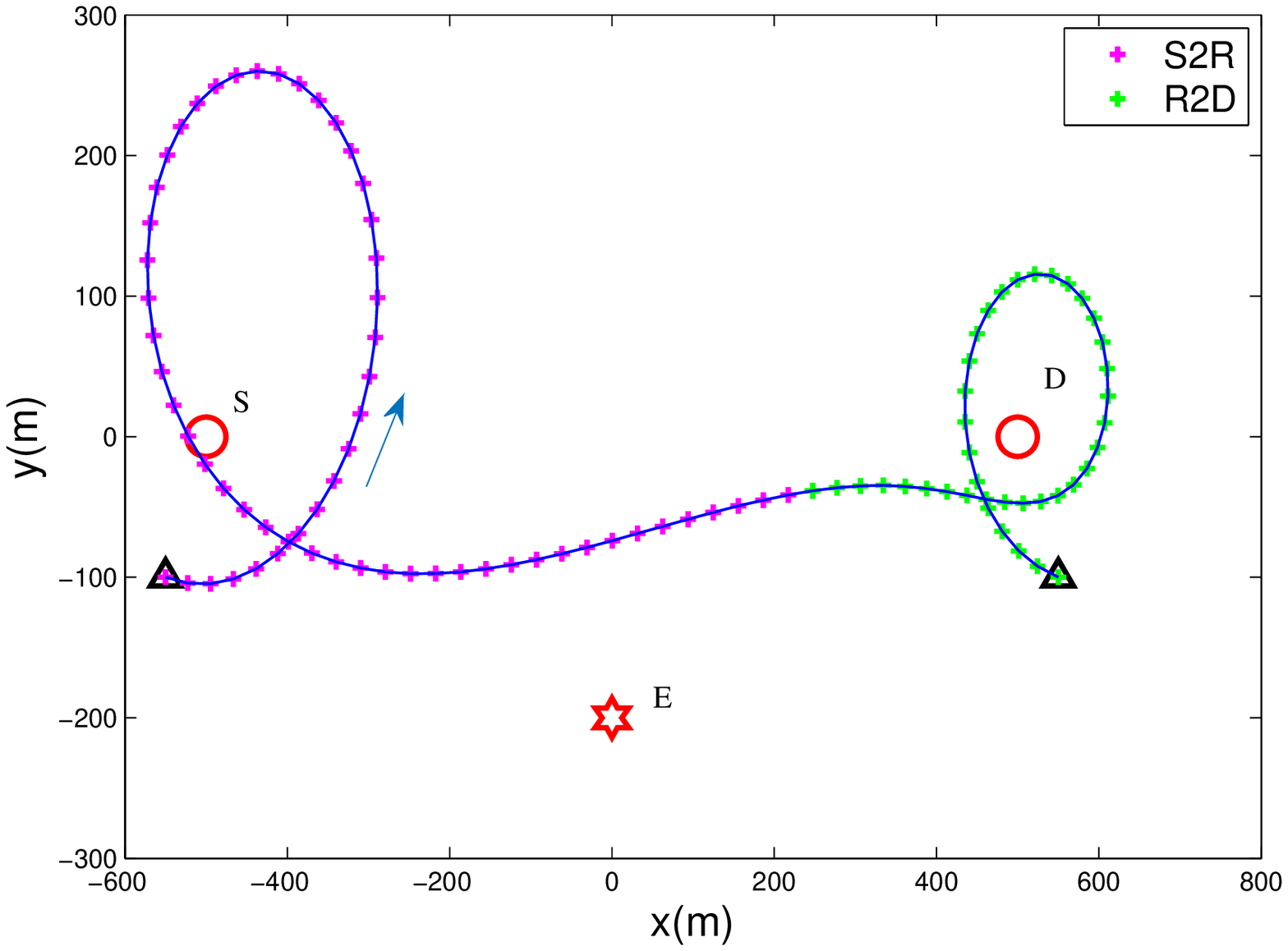}}~~~
\subfigure[$T=150s$]{\includegraphics[width=0.5\textwidth]{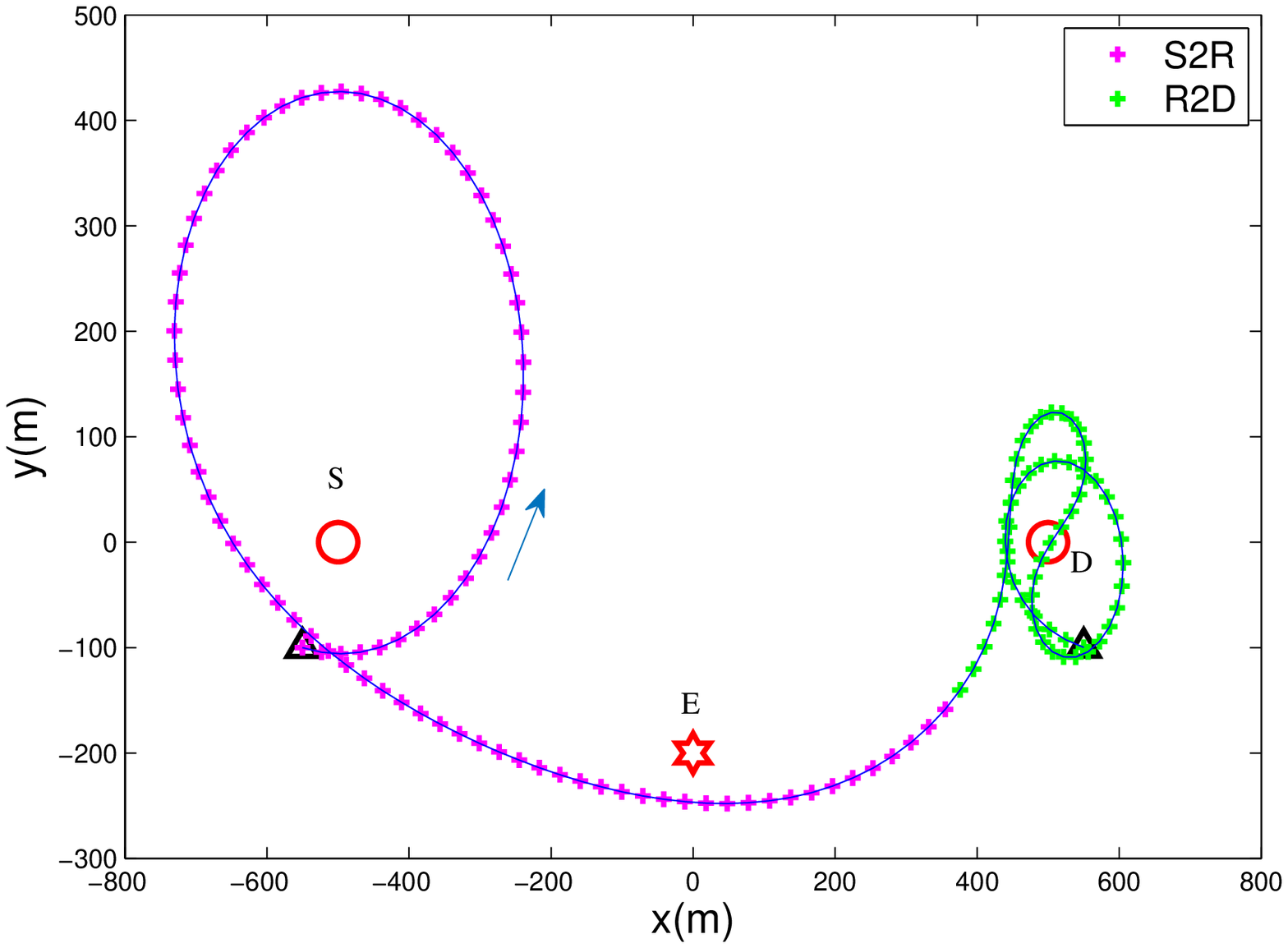}}\\
\subfigure[$T=250s$]{\includegraphics[width=0.5\textwidth]{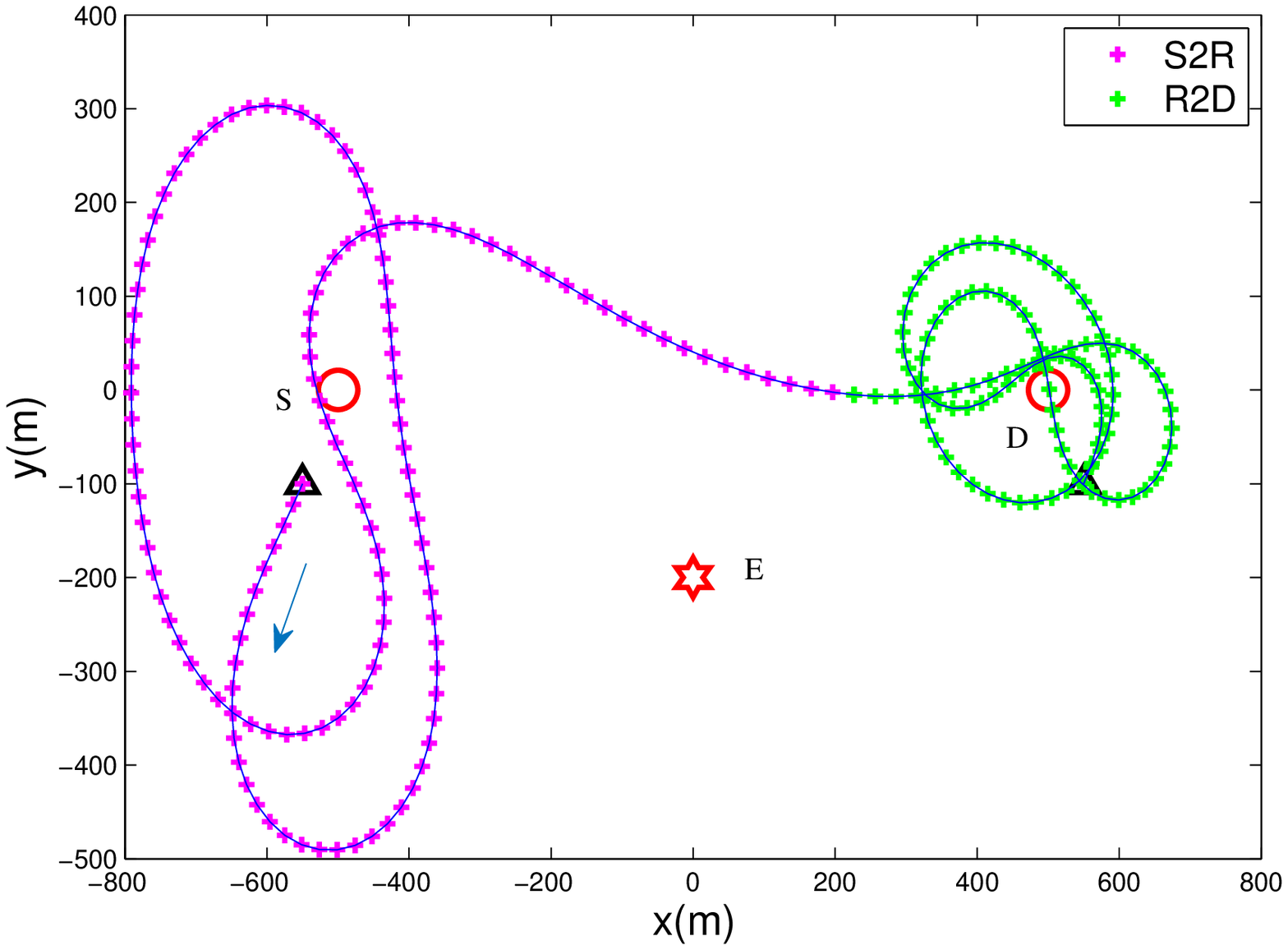}}~~~
\subfigure[$T=300s$]{\includegraphics[width=0.5\textwidth]{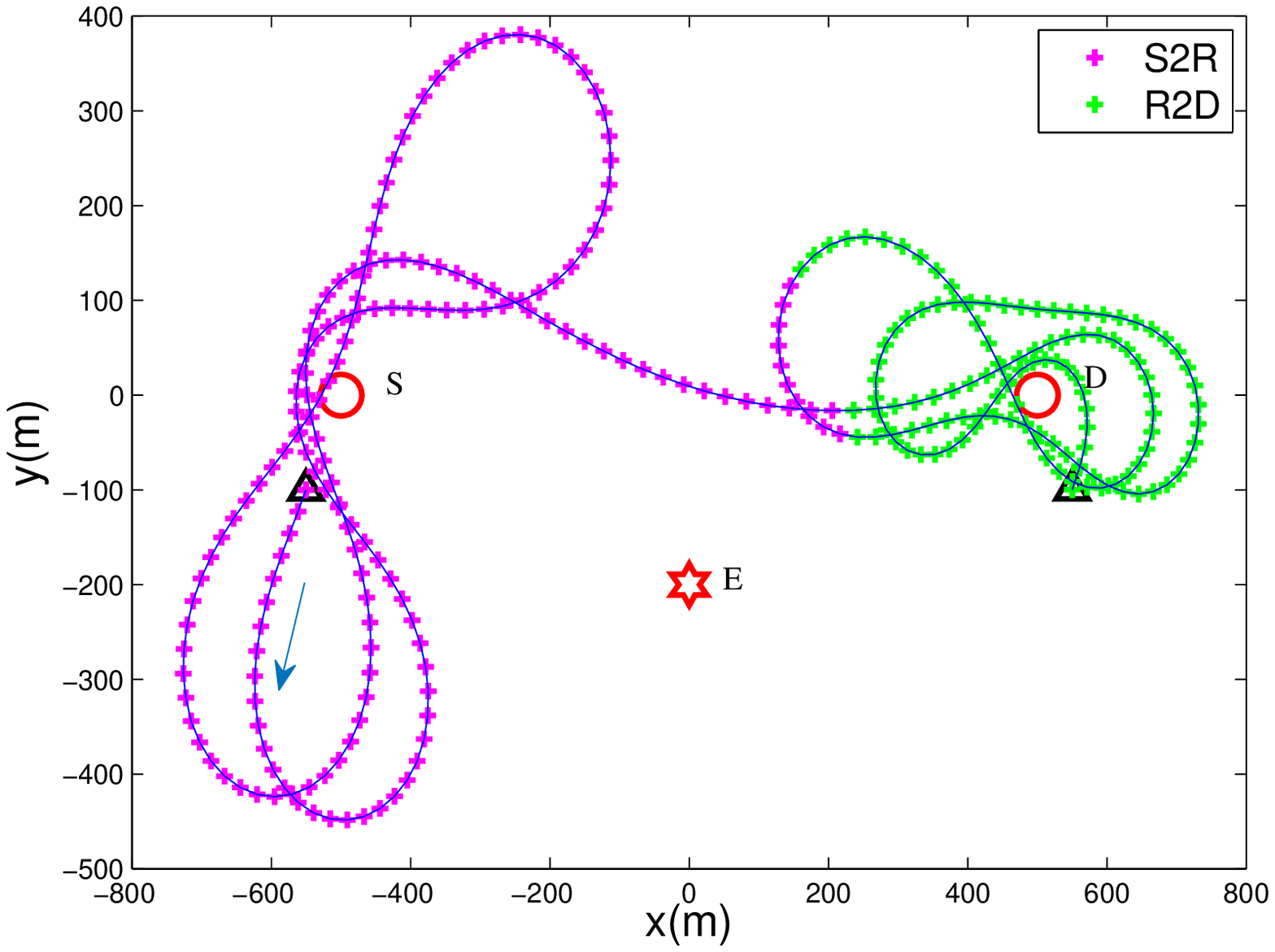}}
\caption{UAV trajectories for different period $T$} \label{UAV_T}%\vspace{-0.3cm}
\end{figure*}
\section{Numerical Results}
In this section, numerical results are provided to validate the proposed algorithm that jointly optimizes the communication scheduling, power allocation and UAV trajectory. Specifically, we adopted the straight flight with optimal resource allocation (denoted by SD/w design) and double-circular flight (denoted by DCF/wo design) as the benchmark compared to the proposed solution.

The locations of source node $S$, legitimate destination node $D$ and eavesdropper node $E$ are fixed at $[-750,0]^T, [750,0]^T$, and $[300,-300]^T$, respectively. The UAV is assume to fly at $H=100m$, and the corresponding initial and final locations is set to $\mathbf{q}_0=[-900,-200]^T$ and $\mathbf{q}_F=[900,-200]^T$, respectively. The communication bandwidth for information reception/transmission is set as $1MHz$. For the ground-to-groud channel between $S$ and $E$, the path loss constant is set as $K=10^{-3}$, and the path loss exponent factor is $\alpha=3$. Moreover, the noise power is set to $\sigma^2=-110dBm$, and the reference channel power is set to $\beta_0=-50dB$. The length of each time slot is set as $\delta_t=1s$. The average transmit power budgets of the source node and UAV is set to $\bar{P}_s=\bar{P}_r=10dBm$. The maximum speed and acceleration of UAV is set as $V_{max}=40M/s$ and $a_{max}=5m/s^2$, respectively. Besides, base on \cite{Zeng2017energyefficient}, we set $c_1=9.26\times10^{-4}$, $c_2=2250$. The accuracy in Algorithm $1$ is set as $\epsilon=10^{-5}$.

Fig.\ref{UAV_T} shows the UAV trajectories for different period $T$. The source node $S$ and legitimate destination node $D$ are marked with $'\circ'$. The eavesdropper node $E$ and the initial/final location are marked with $'\times'$ and $'\triangle'$, respectively. To illustrate the UAV's location and the communication scheduling in each slot, the trajectory is sampled every second with two different colors and the sampled points are marked with $'+'$.

From Fig.\ref{UAV_T}, it can be observed that the UAV first collects information from $S$ (denoted by S2R), then it flies closer to $D$ for data forwarding (denoted by R2D). For a smaller value of $T$ (e.g. $T=100s$), the secrecy rate is increased mainly by optimizing the scheduling factor and power allocation since there is little degree of freedom for trajectory optimization. As $T$ increasing, the trajectory optimization plays an important role in improving the secrecy rate of the system. It is also observed that for a large time horizon $T$, the UAV collects and transmits information in an alternate manner as shown in Fig.\ref{UAV_T}(d). This is because with the increasing of $T$ , the UAV would increase its turning radius to decrease the propulsion energy consumption such that the UAV collects information from $S$ rather than transmit information to $D$ when it flies closer to $E$, which is expected since the objective function of problem $\mathbf{P}1$ is nonnegative. Moreover, in order to improve the secrecy rate of two hop link, the UAV hovers around $S$ and $D$ following an approximately $'8'-shape$ trajectory, as shown in Fig. \ref{UAV_T}(c)-(d).
\begin{figure}
\setlength{\abovecaptionskip}{0.cm}
\setlength{\belowcaptionskip}{-0.cm}
%\begin{minipage}
\centering
{\includegraphics[width=0.5\textwidth,height=2.5in]{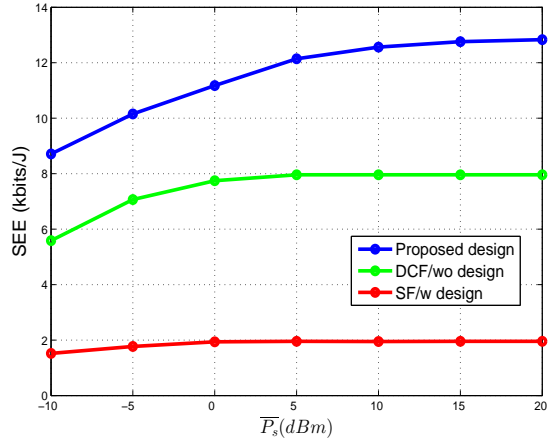}}~~~
\caption{SEE versus average power $\bar{P}_s$ for $T=300s$.\vspace{-3ex}} \label{SEE_P}
\end{figure}
Fig.\ref{SEE_P} shows the secrecy energy efficiency (SEE) of different schemes versus the average power budget $\bar{P}_s$ for time horizon $T=300s$. It is observed that the SEE firstly increases with the increasing of $\bar{P}_s$, as expected. When the value of $\bar{P}_s$ is smaller, the UAV has enough capacity for forwarding the collected secrecy information from $S$ to $D$. As $\bar{P}_s$ increases, the collected information would gradually reach to the upper limit of the UAV's forwarding capacity since $\bar{P}_r$ is limited. Moreover, noted that the proposed method always achieves the best SEE since the UAV has the largest degree of freedom for trajectory optimization.
\begin{figure}
\setlength{\abovecaptionskip}{0.cm}
\setlength{\belowcaptionskip}{-0.cm}
%\begin{minipage}
\centering
{\includegraphics[width=0.5\textwidth,height=2.5in]{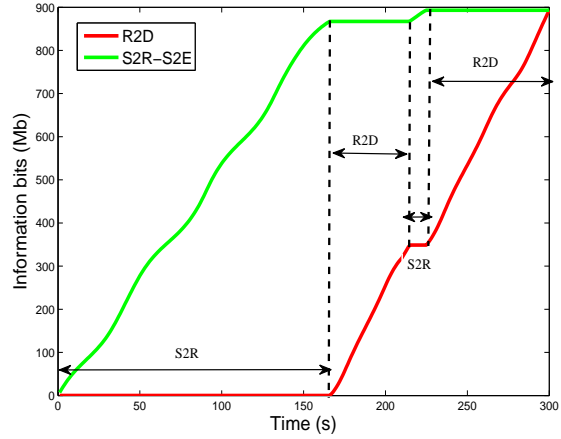}}~~~
\caption{Illustration of the information-causality constraint for $T=300s$.\vspace{-3ex}} \label{causuality}
\end{figure}
In order to show the effectiveness of the information causality in (\ref{casual_c}), the sum of information bits of $S2R$ and the available secrecy bits from $S$ (denote by $S2R-S2E$) versus each time slot for $T=300s$ is plotted in Fig.\ref{causuality}. It can be observed that the whole time horizon is partitioned into four segments due to TDD adopted in the system, which is related to the Fig.\ref{casual_c_lambda}. First, the UAV collects information from $S$, then it forwards the collected information to $D$ when it flies closer to the destination node. Next, since the distance between the UAV and $D$ is larger than that between the UAV and $E$, the UAV chooses to receive the information from S for ensuring the non-zero secrecy rate as shown in (\ref{RD_s_rate}). At the last segment, the UAV transmits the information to $D$ over the optimal communication channels until the information causality constraint in (\ref{casual_c}) holds with equality, which validates our proposed design.
\hspace{-3ex}
\begin{table}
  \centering
  \begin{tabular}{cccc}
\toprule[2pt]
Horizon & Proposed design & DCF/wo design & SF/w design\\
$T(s)$ &  $(kbits/J)$  &  $(kbits/J)$ &  $(kbits/J)$ \\
\midrule[1pt]
\specialrule{0em}{2pt}{2pt}
100 & 12.27 & 5.52 & 3.87\\
\specialrule{0em}{2pt}{2pt}
150 & 12.79 & 8.07 & 2.60\\
\specialrule{0em}{2pt}{2pt}
200 & 12.21 & 7.96 & 1.95\\
\specialrule{0em}{2pt}{2pt}
250 & 12.02 & 7.17 & 1.56\\
\specialrule{0em}{2pt}{2pt}
300 & 11.08 & 6.31 & 1.30\\
\specialrule{0em}{2pt}{2pt}
350 & 10.52 & 5.51 & 1.12\\
\specialrule{0em}{2pt}{2pt}
400 & 11.34 & 4.82 & 0.96\\
\specialrule{0em}{2pt}{2pt}
450 & 10.89 & 4.32 & 0.87\\
\specialrule{0em}{2pt}{2pt}
\bottomrule[2pt]
\specialrule{0em}{2pt}{2pt}
\end{tabular}
  \caption{SEE Comparison of three methods }\label{SEE_comparsion}
\end{table}
Table \ref{SEE_comparsion} shows the SEE comparison between the proposed solution and the benchmarks in different time horizon $T$. It can be observed that the proposed method outperforms the benchmarks for any $T$, which validates that the UAV trajectory optimization plays an important role in SEE improvement, as expected. Through the $T$ increasing,  In addition, the SEE of $SF/w$ design is always lowest since the straight flight constraint gives little freedom for trajectory optimization.
\section{Conclusions}
In this paper, we investigate the physical-layer secure communication in a new UAV-enable mobile relaying system. The objective is to maximize the secrecy energy efficiency of the UAV over a finite time horizon by jointly designing the communication scheduling, transmit power allocation and UAV trajectory, subject to the maximum speed/acceleration and average transmit power constraints. To solve the formulated mixed integer and non-convex problem, an efficient iterative algorithm is proposed by applying SCA method and Dinkelbach’s algorithm. In particular, we adopt two special cases as benchmarks to illustrate the performance of the proposed design. Numerical results validate our proposed algorithm and also show that the SEE of the UAV can be significantly enhanced under the proposed design, as compared to the benchmarks.

\end{document}